\documentclass[fleqn,usenatbib]{mnras}

\usepackage{newtxtext,newtxmath}

\usepackage[T1]{fontenc}

\DeclareRobustCommand{\VAN}[3]{#2}
\let\VANthebibliography\thebibliography
\def\thebibliography{\DeclareRobustCommand{\VAN}[3]{##3}\VANthebibliography}

\usepackage{graphicx}	
\usepackage{amsmath}	
\usepackage[pagewise]{lineno}     
\title[Power pattern uncertainty in 21 cm experiment]{The effects of the antenna power pattern uncertainty within a global 21 cm experiment}

\author[J. Cumner et al.]{
John Cumner,$^{1,2}$\thanks{E-mail: jmc227@cam.ac.uk}
Carla Pieterse,$^{3}$
Dirk de Villiers$^{3}$
and Eloy de Lera Acedo$^{1,2}$
\\
$^{1}$Cavendish Astrophysics, University of Cambridge, Cambridge, UK\\
$^{2}$Kavli Institute for Cosmology in Cambridge, University of Cambridge, Cambridge, UK\\
$^{3}$Department of Electrical and Electronic Engineering, Stellenbosch University, Stellenbosch, South Africa\\
}

\date{Accepted XXX. Received YYY; in original form ZZZ}

\pubyear{2023}

\begin{document}
\label{firstpage}
\pagerange{\pageref{firstpage}--\pageref{lastpage}}
\maketitle

\begin{abstract}
Measuring the redshifted sky-averaged neutral hydrogen 21~cm signal with a wideband antenna operating at metre wavelengths can probe the thermal history of the Universe and the first star and galaxy formation during the Cosmic Dawn. Measurement of this “global 21~cm” signal is extremely challenging due to foreground signals that are orders of magnitude brighter than the cosmological signal, which must be modelled and removed first. The Radio Experiment for the Analysis of Cosmic Hydrogen (REACH) aims to improve this process by simultaneously fitting the full posterior distribution of both the cosmological and foreground signals with Bayesian inference. The method, however, relies
on an informed prior; partially derived from a simulated antenna power pattern. This simulated antenna power pattern will differ from the true antenna power pattern of the deployed instrument, and the impact of this uncertainty is unknown. We investigate this problem by forward modelling mock data with different levels of power pattern uncertainty through
the REACH pipeline. We construct perturbed antenna power patterns through truncation of a singular-value-decomposed simulated power pattern; using one to generate mock observation data and the others to inform the prior. The power pattern uncertainty is quantified as $\Delta D$, the absolute mean of the difference between the original and perturbed power patterns. Comparing the evidence and root-mean-square error we find that $\Delta D$ better than $-35$~dB, equivalent to millimetre accuracy in the antennas dimensions, is necessary for confident detection of the global signal. We discuss potential solutions to achieve this high level of accuracy.

\end{abstract}




\section{Introduction}
The Universe evolved from a simple and smooth medium that emerged from the Big Bang to the plethora of galaxies that surround us today. Observations of the early Universe and local structures have had success in increasing our understanding of the formation of structures in our Universe. However, there remains a substantial gap in our understanding concerning the time of the Cosmic Dawn, when the first luminous structures begin forming.
Probing these epochs is at the frontier of enhancing our insight into early cosmic structure formation \citep{Furlanetto:2006}.

The redshifted emission and absorption signal from neutral hydrogen gas surrounding the first star and galaxies is an essential probe for studying this transformative period. Through the Wouthuysen-Field effect the neutral hydrogen within the universe is able to cause radio emission at the 21~cm wavelength, 1420~MHz, through the transition of electrons between different hyperfine energy levels \citep{Wouthuysen1952,field58}. This radiation is redshifted to lower frequencies as the universe ages and its varying intensity provides a trace of universal evolution, the 21~cm signal.
This paper focuses on the sky-averaged, global, spectrum taken of the redshifted 21~cm signal between 50 and 150~MHz. This signal traces the cooling and heating of hydrogen gas until the ultraviolet radiation produced by these luminous objects ionizes the neutral hydrogen gas, causing the 21~cm emission to decrease and eventually vanish. Thus, the signal profile could be used to infer the underlying processes that governed the formation of the first stars and galaxies. 

 Several ongoing experiments aim to measure the global 21~cm signal, including Experiment to Detect the Global EoR Signature (EDGES; \citep{Bowman2007}), Large-aperture Experiment to Detect the Dark Ages (LEDA; \citep{LEDA_Price2017}), Shaped Antenna measurement of the RAdio Spectrum (SARAS; \citep{SARAS_Singh2018}), Sonda Cosmológica de las Islas para la Detección de Hidrógeno Neutro (SCI-HI; \citep{SCI_HI_Voytek2014}), Probing Radio Intensity at High-Z from Marion (PRIZM; \citep{PRIZM_Philip2018}), Broadband Instrument for the Global HydrOgen ReionizatioN Signal (BIGHORNS; \citep{BIGHORNS_Sokolowski2015}), Mapper of the IGM Spin Temperature (MIST; \citep{Monsalve2023}) and{\color{black}, the experiment considered in this paper,} Radio Experiment for the Analysis of Cosmic Hydrogen (REACH; \citep{2022REACH}). 

{\color{black}A common approach to an attempted measurement of the global 21~cm signal is through the use of a single simple antenna. These instruments have very low angular resolution, with a main beam width up to $60^\circ$, so they observe a large portion of the sky at once. For these instruments to successfully measure a global 21~cm signal it is required that the beam pattern is as smooth as possible in both space and frequency to avoid adding unintended structure to a measurement due to coupling with the sky.}

In 2018, EDGES published a possible first measurement of the global 21~cm signal \citep{Bowman2018}. EDGES detected a deep absorption profile sky radio-spectrum centred at 78~MHz, which could indicate exotic cosmology. However, independent studies, \citep{Singh2019, sims20, bevins21}, have shown that the data could be explained by residual instrumental systematic signals, which have not been modelled in the data analysis. 
SARAS's limits published in 2021 suggest that EDGES best-fitting profile is disfavoured with 95.3\% confidence, and thus the signal may not be evidence for non-standard cosmology \citep{Singh2021}.

The uncertainty in the results demonstrates that the signal is still poorly constrained by existing data, a variety of signal models are still plausible, and the timing of cosmic events is poorly understood. 

Global 21~cm signal experiments often attempt to control and simplify instrumental effects by using a single broadband antenna, however observations remain challenging due to the magnitude of the signal.
The synchrotron radiation from the Milky Way that constitutes most of the foreground emission in these observations is $4-5$ orders of magnitudes brighter than the sky-averaged 21~cm signal. Even slight chromaticity in the antenna's far field power pattern (antenna beam) can introduce unwanted frequency structures in the data, which can be misinterpreted as the 21~cm signal when looking at the residuals in the foreground-subtracted spectra \citep{Hills2018, Thyagarajan2016,Singh2019,monsalve19,sims20,bevins21}. 

Often, the antenna beam is assumed to be smooth and achromatic, which would allow the antenna beam and sky power to be easily separated during analysis. Antenna beams are also typically assumed to be known entirely and predictable using Computational Electromagnetics (CEM) simulations. However, this is never completely the case because of the limited accuracy of the simulation, non-zero manufacturing tolerances, and actual material properties differing from material models, in addition to other effects. 

Another challenge is calibrating the frequency response of the radio antenna \citep{Pritchard2010,Rogers12,roque2021}, this is typically done through comparison to multiple known loads preferably measured to high precision concurrently with the observation to produce the required millikelvin accuracy.

The final spectral structure seen within data is heavily dependent on the convolution between sky power and antenna power pattern. If either of these elements is imperfectly modelled, chromatic structure will be seen within the data analysis. As a perfect model of either the sky or the antenna power pattern is not a realistic expectation it is important to understand the tolerance in both models. While some global 21~cm experiments have done characterization studies in detail \citep{Mahesh2021, Raghunathan2021, Spinelli2022}, the antenna power pattern remains a source of uncertainty. This uncertainty imprints additional unknown spectral features on the sky temperature, ground emission, and receiver noise. If the antenna power pattern is not well known, it can be challenging to distinguish any systematic uncertainties arising from the antenna power pattern or other noise sources compared to a possible global 21~cm signal, which then can compromise detections.

This paper uses the REACH experiment as an example case study. REACH is a novel global 21~cm experiment currently in its commissioning phase in the RFI-quiet Karoo Radio-Astronomy Reserve in South Africa. This experiment aims to complete current observations by tackling the most significant challenge faced by current instruments;  residual systematics. REACH implements Bayesian statistics to analyse and compensate for systematics together with the foreground and the cosmological signal. REACH aims to conduct simultaneous observations with two antennas in the $50 - 175$~MHz band ($z  \sim 7.5-28 $) corresponding to the Cosmic Dawn and Epoch of Reionization, EoR. The REACH antennas were designed through a quantitative figure of merit based approach to be highly synergistic with the Bayesian data analysis pipeline \citep{Anstey2021} developed for the project, thereby providing the best possibility of a confident detection of the global 21~cm signal \citep{anstey2021a,Anstey2023,cumner22}. The antennas are also chosen to be complementary antenna systems, allowing sensitivity to both linearly and circularly polarized radiation.
One is a conical log-spiral antenna, and the other is a hexagonal dipole placed on a 20~x~20~m metallic ground plane with serrated edges to minimize edge reflection.

For this paper, we use the REACH data analysis pipeline to assess the accuracy at which an observing antenna power pattern is required to be known to perform a successful detection of the global 21~cm signal. The REACH data analysis pipeline is a PolyChord \citep{Handley2015a,Handley2015b} based python code that incorporates the effect of beam chromaticity coupled with a non-trivial scaling in frequency of the foreground \citep{Anstey2021,Anstey2023}. This strategy, although computationally costly, leads to robust 21~cm global signal extraction in simulations. A similar method is used in \citep{Sims2023} to compensate for chromaticity within the EDGES antenna system. Other techniques that incorporate beam effects in the foreground model using machine learning methods have been proposed by \citep{Tauscher2021}.

The focus of this paper will be establishing an allowable uncertainty between two normalized antenna power patterns within the REACH data analysis pipeline \citep{Anstey2021}, with one power pattern used to generate input data and the other used within the analysis stage.

First, we introduce some likely causes of antenna beam uncertainty to be present within a global 21~cm experiment in Section~\ref{sec: Uncertianty}. Thereafter, in Section~\ref{sec: definitions}, we define the equations we used to calculate the antenna temperature and antenna power pattern uncertainties. We also discuss the modified REACH dipole used to produce the far field pattern, from which we derive the power patterns used in this paper. We then discuss how the REACH Bayesian data analysis pipeline was set up to calculate the confidence of signal detection. In Section~\ref{sec: Directivity reconstruction} we discuss a selection of methods for reconstructing an antenna power pattern using weighted basis functions to produce perturbed beam patterns. Within section \ref{sec: directivity accuracy} we show how the accuracy of the reconstructed power patterns is quantified and how this relates to the power pattern and antenna temperature uncertainty. In Section~\ref{sec: Errors}, we analyse the impact of beam uncertainty on the confidence and accuracy of detections of various candidate global 21~cm signals using the REACH Bayesian pipeline. The Section~\ref{sec: Power pattern difference} briefly discusses the power pattern difference when physical dimensions of the antenna are varied. In Section~\ref{sec: Compensation} we provide a summary of results with respect to a tolerable level of uncertainty in knowledge of the antenna power pattern in addition to discussing some possible methods for compensating for this uncertainty. 

\section{Possible causes of uncertainties in an antenna beam} \label{sec: Uncertianty}

In order to accurately control and compensate for systematic instrumental effects within a global 21~cm experiment, it is required to have a detailed and accurate understanding of the antenna power pattern, which observes the sky. Several factors can contribute to uncertainty in the antenna power pattern thereby increasing the effective chromaticity of the antenna. These effects are generally not sufficiently well understood as to allow for complete modelling, and hence they remain `unknown' systematics. This includes, but is not limited to:

\begin{enumerate}
    \item The unknown properties of the soil under the antenna. Soil is inhomogeneous, and the complex permittivity can vary with frequency, moisture content, salinity and temperature, making modelling the soil challenging. The effects of the soil can be noticeable on LEDA's signal recovery, as shown in \citep{Spinelli2022}, the size of the ground plane did not seem to mitigate this effect.
    \item The effects of the ground plane's geometry. Ground plane geometry can increase the complexity of the frequency structure of the antenna beam by imposing a frequency ripple. The ripples have lower amplitude but faster oscillation for larger ground planes. \citep{Bradley2019} investigated how EDGES ground plane resonance can result in a systematic artefact producing a broad absorption feature in the spectra. Serrated ground planes are used with the assumption that they will reduce the edge effect of the ground plane. However, it could add more spectral complexity to the antenna beam, which is hard to model. \citep{Spinelli2022} showed that the serrations might increase the sensitivity of the antenna beam to the soil properties.
    \item Geometric perturbations of the physical structure. The geometry of mechanical structures can differ from the simulated antenna geometry due to mechanical, manufacturing and construction tolerances. Geometry can also be time-varying due to mechanical deflection, wear and thermal expansion. All of these will cause perturbations in the beam pattern.
    \item Uncertainty in the accuracy of antenna simulations. CEM simulations have finite accuracy, mostly due to the inherent differences between the model and physical structure. Results might also differ between different software and solvers. Numerical errors can also be introduced due to pixelization and discretization of the model \citep{Mahesh2021}.
   \item The horizon and other structures in the environment might be unknown and uncompensated for \citep{Pattison2023}. Horizon effects have both time and spectral dependencies and at low frequencies \citep{Bassett2021} have shown that these effects cannot be ignored. Bias can be introduced by ignoring attenuation produced by surrounding foreground obstructions such as man-made structures, terrain and vegetation. 
\end{enumerate}

It becomes clear that there are abundant sources of uncertainty within the radiometer setup from both the physical structure itself and within the environment. Some of these uncertainties are also time-varying. It is thus impossible to completely model, understand and compensate for all systematics. It is, however, essential to understand and try to quantify how much our presumed `known' antenna beam needs to correspond with the actual antenna beam of the observing instrument in the field and how this relates to how confident one can be about detecting the global 21~cm signal and not preventing a detection through the presence of a systematic uncertainty within the system.

\section{Methodology}
\label{sec: definitions}
In this section, we introduce antenna power patterns and antenna temperature. We allow the quantification of uncertainty in the antenna beam by defining equations that describe the difference between a pair of antenna power patterns and similarly for a pair of antenna temperatures. A modified REACH dipole is introduced, and its simulated antenna beam is used as the base antenna beam for the remainder of this paper. Finally, we describe the REACH Bayesian data analysis pipeline which is used to assess the capability of various power patterns to detect the global 21~cm signal. 

  \subsection{Normalized antenna power pattern}
    We shall describe the antenna beam in terms of its normalized power pattern, $D(\nu,\Omega)$, as
    \begin{equation}
        D(\nu,\Omega)\propto \left|E_\text{1}(\nu,\Omega)\right|^2 + \left|E_\text{2}(\nu,\Omega)\right|^2  ,
    \end{equation}
  where  $E_1(\nu,\Omega)$ and $E_{2}(\nu,\Omega)$ are the orthogonal set of the co- and cross-polarised electric fields that are variable over frequency, $\nu$ and angular direction $\Omega$. These E-fields represent the spacial pattern of the received E-field strength of radiation by the antenna over direction and frequency. They therefore describe the coupling between a passively observing antenna and the sky power. Although it is likely that the patterns of the differing polarizations will be different the effects of polarization are outside the scope of this work and we shall instead examine the effects on the gain pattern,
\begin{equation}
    D(\nu,\Omega) = \frac{4\pi\left|\left|E_\text{1}(\nu,\Omega)\right|^2 + \left|E_\text{2}(\nu,\Omega)\right|^2\right|}{\int_{4\pi}\left|E_\text{1}(\nu,\Omega)\right|^2 + \left|E_\text{2}(\nu,\Omega)\right|^2 d\Omega}.
    \label{eqn: DfromE}
\end{equation}
 We shall enforce,
    \begin{equation}
        D(\nu,\theta > \pi/2) = 0,
    \end{equation} 
    where $\theta$ is the zenith angle, to remove the requirement to model the near-field contribution of the soil to the antenna temperature.
  Further, we enforce the normalization of the power  patterns such that, 
    \begin{equation}
    \int_{4\pi} D(\nu_i,\Omega) d\Omega = 4\pi,
    \label{eqn: directivity norm}
    \end{equation}
    at each frequency, $\nu_i$.

  For the purposes of this paper it is important to quantify the difference between two power patterns with a single number. To do this we shall define the metric
  \begin{equation}
        \Delta D = \langle\langle |D(\nu,\Omega) - \Tilde{D}(\nu,\Omega)|\rangle_\Omega\rangle_\nu,
        \label{eqn: delta D}
    \end{equation}
    where $\langle\rangle_{\Omega}$ denotes the mean average over all angles, and $\langle\rangle_{\nu}$ the mean average over all frequencies. For the purposes of this paper we shall use $D(\nu,\Omega)$ to denote a base power pattern, and $\Tilde{D}(\nu,\Omega)$ to denote reconstructed, effectively perturbed, power patterns.

    This metric averages out the fine detail information pertaining to spacial and frequency variation between two power  patterns, but acts as a representative single value guide to the difference between them. For the remainder of this work, we shall use $\Delta D$ in dB, with larger negative values indicating a closer agreement between $D(\nu,\Omega)$ and $\Tilde{D}(\nu,\Omega)$.

\subsection{Antenna temperature} \label{subsec: TA}  
For the case of global 21~cm experiments, it is customary to refer to antenna temperature rather than power when considering the output of an observing antenna. The general form of the antenna temperature is calculated as 
    \begin{equation}
   T_{\text{a}}(\nu,t) = \frac{\int_{4\pi} D(\nu,\Omega)\eta(\nu)T_{\text{sky}}(t,\nu,\Omega)d\Omega}{\int_{4\pi} D(\nu,\Omega)d\Omega},
\label{eqn: T_ant_t} 
\end{equation}
where $\eta(\nu)$ is the radiation efficiency of the antenna, assumed to be 1 for this paper; $T_{\text{sky}}(t,\nu,\Omega)$ is the sky temperature, variant over time, $t$. Due to the normalization of $D(\nu,\Omega)$ enforced by \eqref{eqn: directivity norm}, $4\pi$ can be used instead as the denominator.  This convention will be followed throughout this paper.

$T_{\text{a}}(\nu,t)$ is integrated over some time interval $\tau$ to give,
\begin{equation}
    \label{eqn: T_A}
    T_\text{A}(\nu) = \int_\tau T_\text{a}(\nu,t) dt.
\end{equation}
For the simulated observation of the sky a scaled version of the 408~MHz Haslam all sky survey \citep{Haslam1982,Remazeilles2015} is used for the sky temperature distribution $T_\text{sky 408}$. To extrapolate this sky map to the relevant 50~-~130~MHz bandwidth for this paper a synchrotron like $-2.5$ index power law is used, with the cosmic microwave background temperature $T_\text{CMB} = 2.725$~K to give
\begin{equation}
T_{\text{sky}} (\nu,\Omega) = \left( T_{\text{sky 408}}(\Omega) - T_{\text{CMB}} \right) \left(\frac{\nu}{408~\text{MHz}}\right)^{-2.5} + T_{\text{CMB}}.
\label{eqn: T_sky}
\end{equation}
We assume the antenna is located at the REACH site in the Karoo Radio-Astronomy Reserve South Africa, $(21.45^\circ, -30.71^\circ)$. Snapshot antenna temperatures are taken with the galaxy above the horizon at {\color{black}LST = 16.5 hour angle} 
($T_\text{sky-gal-up}$) and with the galaxy below the horizon at {\color{black}LST = 22.5 hour angle} 
($T_\text{sky-gal-down}$). Multi-time observations, including those in Section~\ref{sec: Errors}, are started at {\color{black}LST = 4.64 hour angle at the REACH site with a duration of 4 hours, the default time settings for the REACH pipeline} which encompasses the galaxy rising and then falling within the sky over the night.

\begin{figure}
    \centering
    \includegraphics[width = \linewidth]{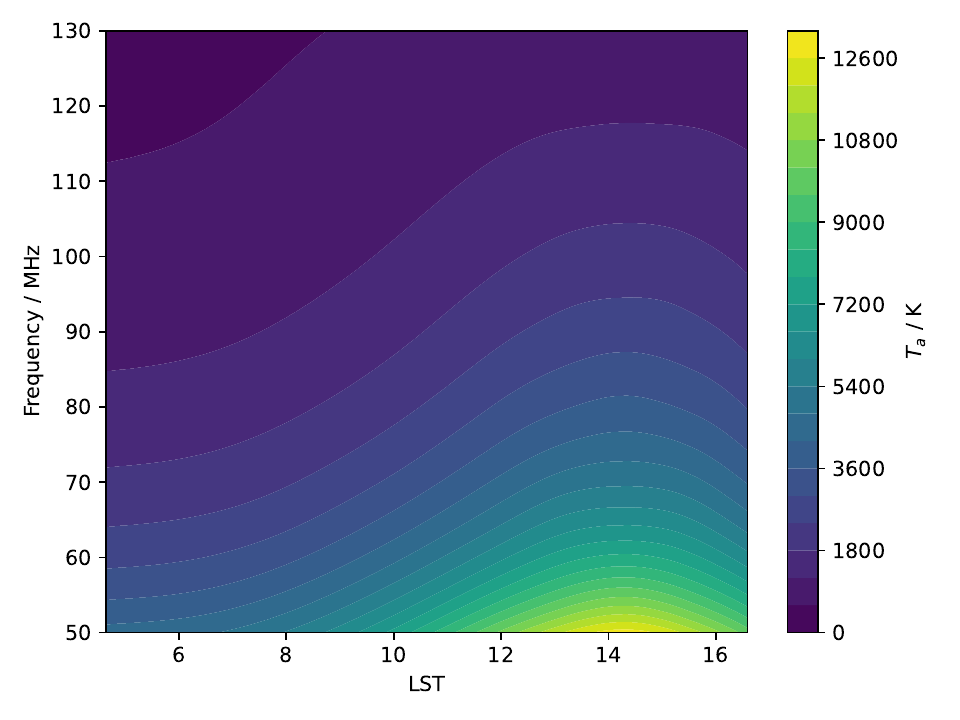}
    \caption{Waterfall plot of the antenna temperature for the modified REACH dipole over a 12 hour period at the REACH site starting.}
    \label{fig: waterfall plot}
\end{figure}

By default, $T_\text{sky}$ is assumed not to include the global 21~cm signal. Which can then be added after the scaling of the sky map as,
\begin{equation}
    \label{eqn T_sky 21}
    T_\text{sky-21}(\nu,\Omega) = T_\text{sky}(\nu,\Omega)+T_{21}(\nu).
\end{equation}
An example waterfall plot of the antenna temperature is shown in Figure~\ref{fig: waterfall plot}, demonstrating the wide range of system temperatures seen over both frequency and time, ranging over 12000~K with frequency and up to 5000~K over time at low frequencies as the galaxy passes through the view.

Time separated bins are used within the simulated observation to improve the accuracy of the REACH data analysis pipeline \citep{Anstey2021,Anstey2023}; particularly for fitting signals at low frequencies. This method will also be used during actual observations where data will be stored in short, time averaged bins.

To calculate the system temperature for an observing antenna, it is required to modify $T_\text{a}$ with the reflection coefficient of the antenna in addition to system gain and additional noise. For this work these effects are neglected, as they occur after the reception of the sky radiation and so will be impartial to the uncertainties within the antenna's beam. During the data analysis these effects are expected to be calibrated out \citep{roque2021}, we shall assume this is the case for this paper.

\subsubsection{Antenna temperature uncertainty}
The expected magnitude of the $T_\text{21}$ is 500~mK or smaller, so an absolute antenna temperature uncertainty between two antenna temperatures, $T_{a}$ and $\Tilde{T}_{a}$, below this value is desirable to avoid obscuring a possible detection of the global 21~cm signal,
    \begin{equation}
        \label{eqn: abs DTa}
        |T_{\text{a}} - \Tilde{T}_{\text{a}}| \ll 500 \text{ mK}~\forall~\nu, \Omega, t.
    \end{equation}
Due to the large variation seen within the overall sky temperature over both time and frequency this absolute value of antenna temperature difference is not ideal for general comparison between two beam patterns. For this reason it is useful to define a fractional difference to allow comparison over frequency and observation time as
    \begin{equation}
        \label{eqn: DT}
        \Delta T_\text{a}/T = \frac{|T_{\text{a}}(\nu,t) - \Tilde{T}_{\text{a}}(\nu,t)|}{T_{\text{a}}(\nu,t)}.
    \end{equation}
For the comparison between the $T_\text{a}$ for different power  patterns we shall use averaging over frequency and the two extreme sky maps, $T_\text{sky-gal-down}$ and $T_\text{sky-gal-down}$. This gives a good overview of the likely $T_\text{a}$ values without the requirement to compute over many time bins. For the impact of a longer observation with the galaxy moving over the sky, a 12~hour computation using 144 time bins is also used from 4.64 hour angle LST. To compact the 12~hour sweep to a single number an average over all frequency and time can be taken. These calculations are undertaken using \textit{HEALpix}\footnote{http://healpix.sourceforge.net} grid \citep{Gorski2005_Healpix} $N_\text{side}=512$ maps, {\color{black}maximum pixel diameter of $0.22^\circ$}, to match the resolution of the sky map, which requires interpolation of the power  pattern from the $1^\circ$ accuracy calculated within \textit{CST microwave studio} \citep{cst}. This is done using a combination of SciPy's griddata linear interpolation and healpy's coordinate to pixel functions.

\subsection{The modified REACH dipole}
\label{subsec: mod dipole}
\begin{figure}
    \centering
    \includegraphics[width=\linewidth]{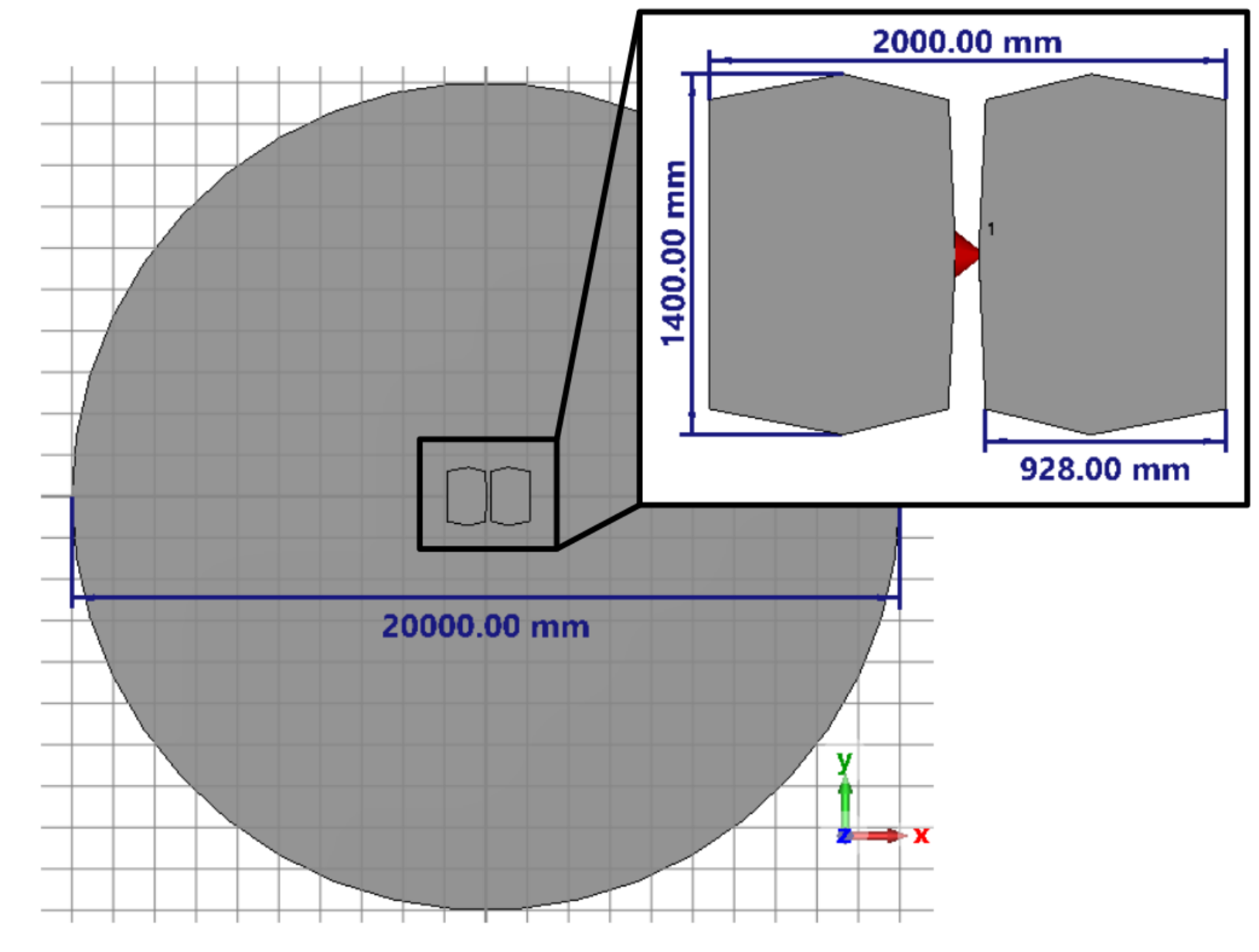}
    \caption{The CST model for the modified REACH dipole antenna used for the base power  pattern throughout this paper. The blade dipole is activated using a discrete port and placed 1~m above a 10~m radius circular ground plane.}
    \label{fig: modified dipole}
\end{figure}
This paper employs a simplified version of the REACH hexagonal blade dipole antenna, shown in Figure~\ref{fig: modified dipole} \citep{cumner22, 2022REACH}, as an example case to analyse the possible effects of antenna beam uncertainty. The antenna is simplified to focus on the primary far field pattern of the antenna without any other structures that cause additional spectral complexity. The balun is removed as impedance is not considered for this work, and a discrete port between the central point of the blades excites the system. A 10~m radius circular ground plane is used for the simulations in order to allow for up to a 50\% reduction in computation time for the CEM model evaluation. The dipole is operated over a 50~MHz to 130~MHz bandwidth, and the antenna structure was meshed using an adaptive hexahedral mesh and simulated in CST with the time domain solver.

\subsection{REACH Bayesian Pipeline} \label{sec: Pipeline}
\begin{figure}
    \centering
    \includegraphics[width = \linewidth]{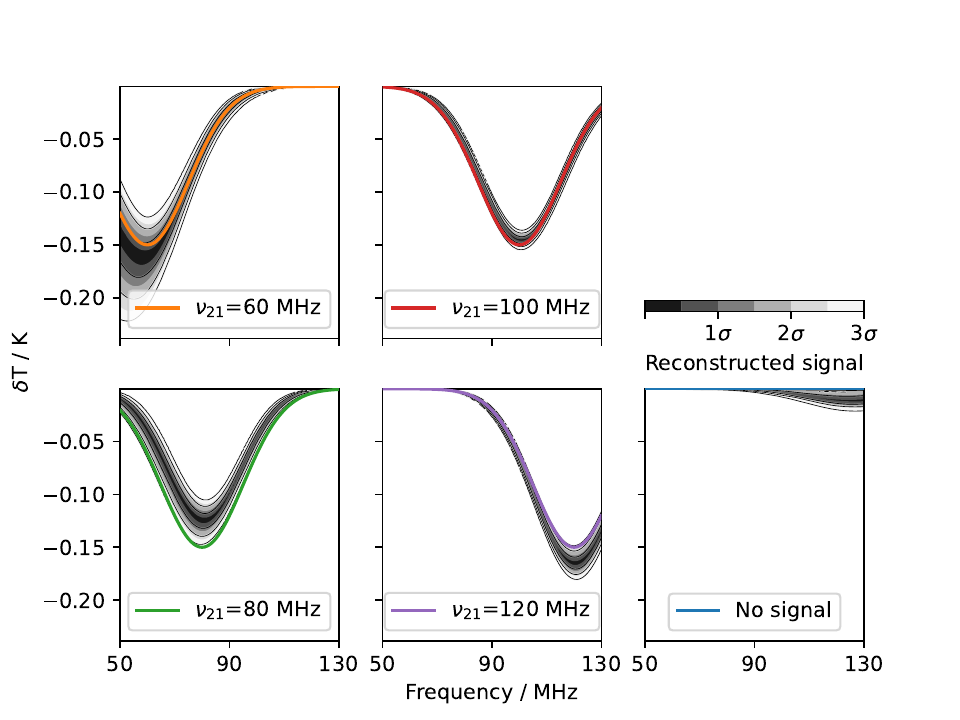}
    \caption{The reconstructed fits using $\Tilde{D}(\nu,\Omega) = D(\nu,\Omega)$, showing the refits for each of the signals in table~\ref{tab: pipeline signals}. Demonstrating that if the power  pattern is known perfectly the REACH pipeline is able to accurately detect the range of input signal examined in this paper.}
    \label{fig: exact beam refit plots}
\end{figure}

The REACH data analysis pipeline \citep{Anstey2021} fits for the spectral indices of a sky map subdivided into regions, with the possibility of including a global 21~cm signal, allowing for an interpretation of the accuracy and confidence of a possible detected signal using Bayesian inference techniques. An example of a resulting fit and residuals are shown in Figure~\ref{fig: exact beam refit plots}.
The confidence in this fit is evaluated with the Bayesian evidence, $\log \mathcal{Z}$. To consider the confidence in any detected signal $\log \mathcal{Z}_{\text{21 cm signal}}$, the pipeline is also run only fitting a foreground model, $\log \mathcal{Z}_{\text{Foreground-only}}$, for each observation data set. The difference between the evidence with just foregrounds fitted and the evidence of the detected signal then gives,
\begin{equation}
    \label{eqn: DLZ}
    \Delta \log \mathcal{Z} = \log \mathcal{Z}_{\text{21 cm signal}} - \log \mathcal{Z}_{\text{Foreground-only}}.
\end{equation}
The value of $\Delta \log \mathcal{Z}$ acts as a measure of confidence in the presence of a signal against none; the higher this number the more confidence of a signal of some type being present within the data. In addition to $\Delta \log \mathcal{Z}$ the root-mean-squared error (RMSE)  between the injected, $T_{\text{G21-input}}$, and reconstructed, $T_{\text{G21-recon}}$, signals is given by 
\begin{equation}
    \label{eqn: RMSE}
    \text{RMSE} = \sqrt{\sum(T_{\text{G21-recon}} -  T_{\text{G21-input}})^2}.
\end{equation}
The RMSE allows for identification of an inaccurate detection, or an uncertain but correctly located detection to be identified. A high $\Delta \log \mathcal{Z}$ indicates a likely signal, and a low RMSE indicates an accurate reconstruction of an injected signal. Considering $\log \mathcal{Z}$ indicates the overall confidence in the fitted parameters compared between fits.

\section{Reconstructing antenna power patterns} \label{sec: Directivity reconstruction}

Here we detail the method by which we construct perturbed power  patterns, $\Tilde{D}(\nu,\Omega)$, of variable uncertainty when compared to a base observing power  pattern, $D(\nu,\Omega)$. This is done through the decomposition of $D(\nu,\Omega)$ into component parts and then reconstructing with limited numbers of these components.

\subsection{Reconstruction using spherical harmonic functions}
\label{sec: sphericals recon}
For a spherical pattern, a straightforward initial exploration into power pattern reconstruction is to use spherical harmonic functions. Typically, a sum of spherical harmonic functions is well suited to describing the behaviour of a function on the surface of a sphere, as we require for a power pattern. So an exploration of the use of the reconstruction,
\begin{equation}
    \Tilde{D}(\nu,\Omega) = \sum_{\ell=0}^p \sum_{m=-\ell}^\ell a_{\ell m}(\nu) Y_{\ell m}(\Omega),
    \label{eqn: sphereical harmonic sum}
\end{equation}
is informative. Here $a_{\ell m}(\nu)$ are frequency varying coefficients of the spacial information contained within the spherical harmonic functions, $Y_{\ell m}(\Omega)$, and the value of $p$ governs the maximum number of coefficients used. The values of $a_{\ell m}(\nu)$ are found using the \textit{map2alm\_lsq} function inbuilt to \textit{healpy}. This method performs an iterative least squares fit for given maximum $\ell$ and $m$ values.

A major drawback of this method is the generic nature of the spherical harmonic basis functions. This will often lead to excess parameters being used to fully describe a given $D(\nu,\Omega)$, which will prevent large scale parameter exploration to be carried out. The reasons for preferring methods that use fewer basis functions are discussed in Section~\ref{sec: Compensation}.

\begin{figure}
    \centering
    \includegraphics[width = \linewidth]{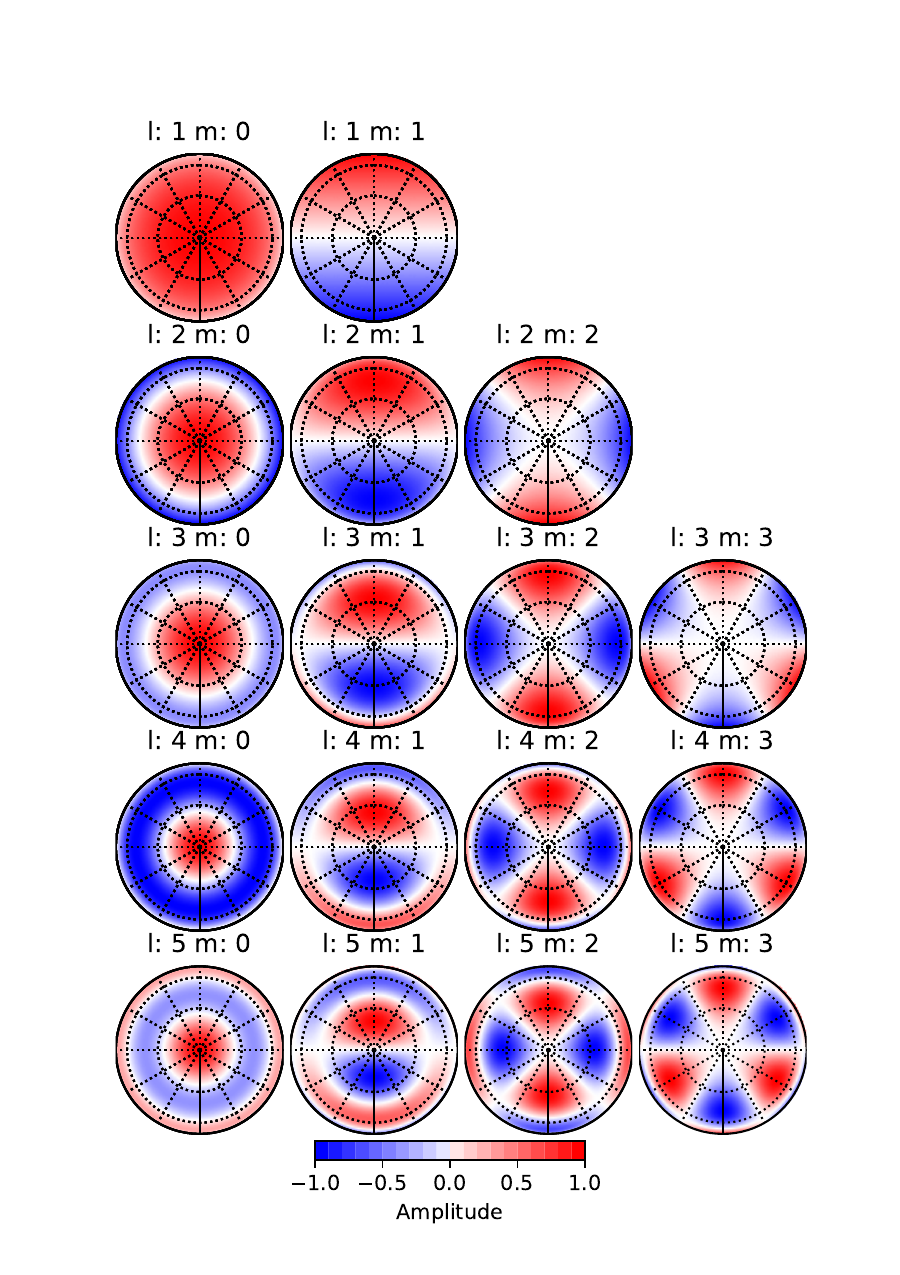}
    \caption{{\color{black}Depictions of the low order spherical harmonic basis functions, which can be used as spacial basis functions.}}
    \label{fig: spherical harmonics}
\end{figure}

\subsection{Reconstruction using numerical basis functions}
\label{sec: SVD}
Instead of using analytical functions, such as spherical harmonics, to reconstruct the physical antenna power pattern, one may also use singular value decomposition (SVD) of the antenna beam pattern to obtain numerical basis functions. As these basis functions are physically motivated, it is expected that fewer functions will be required to accurately reproduce variations within a realistic antenna power pattern. First, we recall the general SVD method for a matrix $\textbf{M}$. When we take the SVD of $\textbf{M}$ we decompose it into three matrices namely the singular values, ($\boldsymbol{\Sigma}$), left singular vectors ($\textbf{U}$) and right singular vectors ($\textbf{V}$) given by equation
\begin{equation}
    \textbf{M} = \textbf{U}\boldsymbol{\Sigma}\textbf{V}^\dag,
    \label{eqn: SVD}
\end{equation}
where $\textbf{U}$ and $\textbf{V}$ give the basis functions of $\textbf{M}$ in row space and column space respectively, and $^\dag$ indicates the conjugate transpose. {\color{black}$\boldsymbol{\Sigma} = \textbf{I}\boldsymbol{\sigma}$} with $\sigma_i$ being the singular values of the decomposition. The magnitude of $\sigma_i$ can be used as an indicator of the magnitude of the contribution of each basis function. This decomposition can either be applied directly to the real power pattern, producing real weights and basis functions, or to the complex electric fields of the pattern producing complex coefficients and basis functions; {\color{black} these two methods are detailed in sections~\ref{sec: PPDM} and \ref{sec: CBFP D} respectively}.

\subsubsection{Power Pattern Decomposition Method, PPDM}
\label{sec: PPDM}
For the power pattern decomposition method (PPDM) $\textbf{M}$ is set to the stacked components of $D(\nu,\Omega)$  and thus $\textbf{M}\in\mathbb{R}^{N_{\nu}~\text{x}~N_\Omega}$ where $N_\nu$ is the number of frequencies sampled and  {\color{black}$N_\Omega$ is the number of spacial angles sampled, corresponding to} the number of pixels in the \textit{HEALpix} map, used to describe the angular component of the power pattern function. 
The stacked power pattern can then be decomposed into $\textbf{U}$, $\textbf{V}$ and $\boldsymbol{\Sigma}$ using \eqref{eqn: SVD}. From these components we then construct weights, $\textbf{W} = \textbf{U}\boldsymbol{\Sigma}$ to encode the frequency dependence of the basis functions and the angular dependence is described through the basis functions $\textbf{V}^\dag$. These components are then used to reconstruct the antenna power pattern as,
\begin{equation}
    \Tilde{D}(\nu,\Omega) = \sum_{i=0}^{p}\text{W}_i(\nu)\text{V}^\dag_i(\Omega).
    \label{eqn: PPDM}
\end{equation}
The value of $p$ is selected such that the value of $\Delta D$ is below a chosen threshold. 
$\Tilde{D}(\nu,\Omega)$, inherits its angular information about the antenna pattern from $\textbf{V}$ and the frequency variation from $\textbf{W}$.

\begin{figure}
    \centering
    \includegraphics[width = \linewidth]{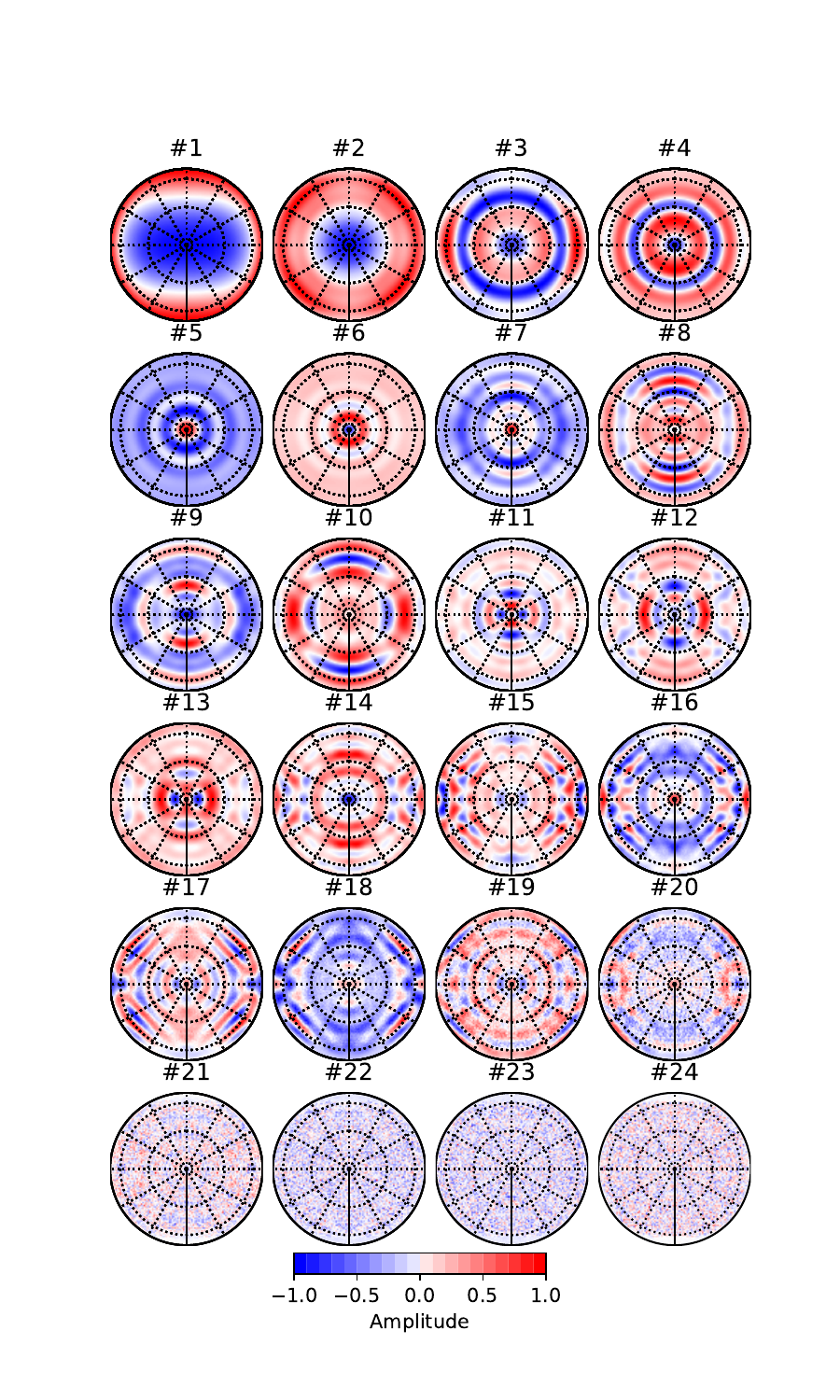}
    \caption{The first 24 PPDM angular basis functions for power pattern construction. \color{black}Constructed from 81 frequency samples from 50 to 130 MHz.}
    \label{fig: first 24 SVD vh}
\end{figure}

 Figure~\ref{fig: first 24 SVD vh} shows the first 24 normalized angular power pattern basis functions, $\textbf{V}^\dag$. They are ordered such that the most significant are the lowest numbered within this plot. The higher number basis functions are seen to become more rapidly varying, until they eventually decay into numerical noise. Inclusion of more than about $20$ basis functions in this case produces diminishing returns, as seen in Figure~\ref{fig: D vs functions}.

\color{black}

\subsubsection{Electric field Decomposition Method, EFDM}
\label{sec: CBFP D}
A similar method, based on expansion of the electric fields instead of the power pattern, has been extensively used in the antenna literature  \citep{Young2013,CBFP_array,Mutonkole2016, Maaskant2012}. Numerical electric field basis functions may be used to reconstruct the full antenna pattern, termed the Characteristic Basis Function Patterns (CBFP) in much of the prior work. 

\begin{figure}
    \centering
    \includegraphics[width = \linewidth]{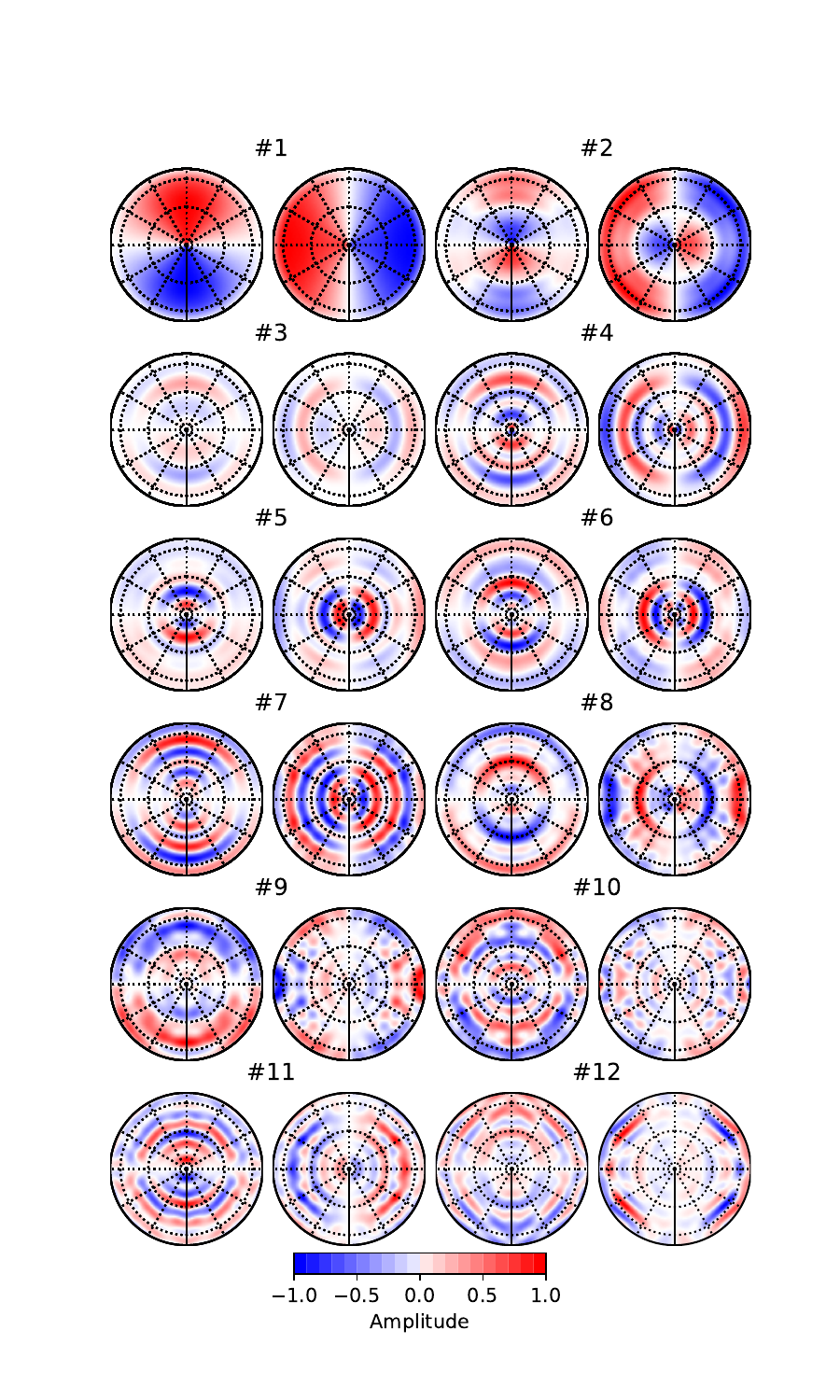}
    \caption{The cross- and co-polar components of the first 12 EFDM angular basis functions. \color{black}Constructed from 81 frequency samples from 50 to 130 MHz.}
    \label{fig: first 24 EFDM}
\end{figure}

 For this method $\textbf{M}$ is the set to column stacked co- and cross-polarised electric fields, $E_1(\nu,\Omega)$ and $E_{2}(\nu,\Omega)$, combined in one matrix $\mathbf{E}$ which is obtained from the CEM simulation as $\textbf{M}\in\mathbb{C}^{2N_\Omega ~\text{x}~N_{\nu}}$ where $N_\Omega$ is the number of angles the electric field is sampled at. $\textbf{U}$ contains the angular basis functions of the electric fields.
The complex weights of the basis functions, $\mathbf{W}$ are calculated by
 \begin{equation}
    \mathbf{W} = \textbf{U}^{\dag}\mathbf{E},
    \label{eqn:W CBFP} 
\end{equation}
which results in the equivalent to \eqref{eqn: PPDM},
\begin{equation}
    \Tilde{E}(\nu, \Omega) = \sum_{i=0}^{p}\text{U}_i(\Omega)\text{W}_i(\nu).
    \label{eqn: EFDM}
\end{equation}
$\Tilde{D}(\nu,\Omega)$ may hence be reconstructed from the $\Tilde{E}(\Omega,\nu)$ using \eqref{eqn: DfromE}. Figure~\ref{fig: first 24 EFDM} shows the cross- and co-polar components of the EFDM normalized angular basis functions with the greatest contribution to the rebuilt power patterns.

\section{Decomposition results}\label{sec: directivity accuracy}
In this section, we shall discuss how the number of basis functions used in the methods in Section~\ref{sec: Directivity reconstruction} relates to $\Delta D$. Also, we relate $\Delta D$ to $\Delta T_{a}/T$ for when the galaxy is in full view of the beam and when the galaxy is down.

\subsection{Spherical harmonic decomposition}
Table~\ref{tab: spherical harmoinics accuracy} shows the accuracy, $\Delta D$, achieved for various numbers of spherical harmonics coefficients. We see that $p=9$ requiring 55 coefficients yields $\Delta D = -10.01\text{ dB}$. To reach $\Delta D \leq -30\text{ dB}$ requires $p=29$ with 465 coefficients. Clearly more harmonics improve the reconstruction, but accurate reconstruction needs very large number of coefficients for such electrically large antennas. {\color{black} Shown in Figure~\ref{fig: spherical harmonics} are the first 17 spacially variant spherical harmonic basis functions, comparing these to the basis functions shown in Figure~\ref{fig: first 24 SVD vh} and \ref{fig: first 24 EFDM} shows the lack of resolution present within the analytical functions compared to those constructed using SVD methods. The simplicity of the spacial basis functions in this method leads to a reasonable first exploration, however greater accuracy can be achieved more efficiently with the SVD-based basis functions, the accuracy of which will be discussed next.}
\begin{table}
    \centering
    \caption{The number $\Delta D$ produced for different numbers of spherical harmonic coefficients.}
    \begin{tabular}{lcr}
    \hline
    p value & \# coefficients & $\Delta D$ \\
    \hline
    1  & 3  & 2.1  \\
    4  & 15  & -5.96  \\
    9  & 55  & -10.01  \\
    16  & 153  & -15.03  \\
    22  & 276  & -20.52  \\
    29  & 465  & -30.47  \\
    44  & 1035  & -35.09  \\
    \hline
    \end{tabular}
    \label{tab: spherical harmoinics accuracy}
\end{table}

 \subsection{Singular value decomposition methods}
\begin{figure}
        \centering
        \includegraphics[width = \linewidth]{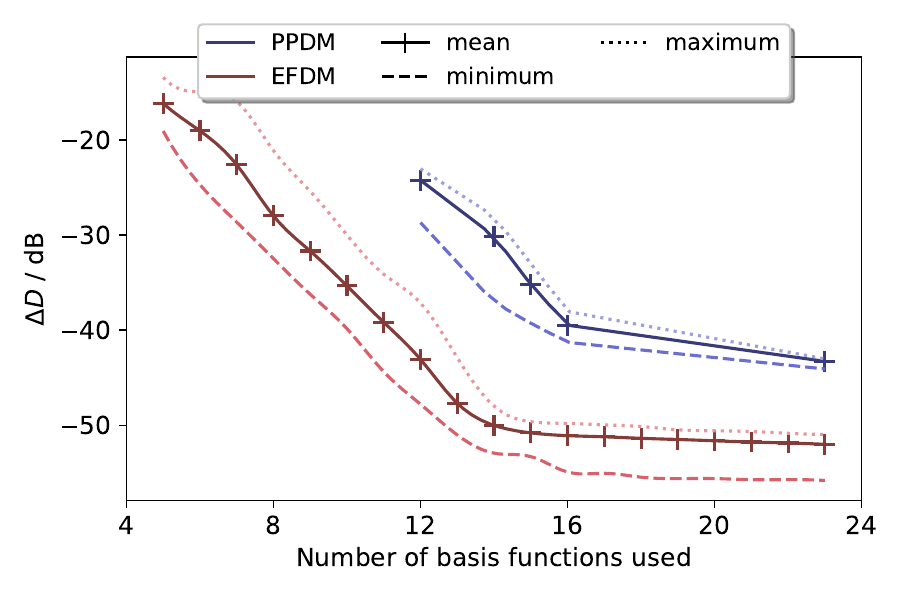}
        \caption{Comparison between $\Delta D$ and number of basis functions used for both the PPDM and EFDM reconstructions. Showing the improved performance per coefficient of the EFDM along with the turning point at which improved accuracy is no longer seen for increased numbers of coefficients.}
        \label{fig: D vs functions}
    \end{figure}

 Using both the PPDM and EFDM for reconstructing $\Tilde{D}(\nu,\Omega)$, we are able to produce strong relationships between the number of basis functions and $\Delta D$. Figure~\ref{fig: D vs functions} shows the distributions for accuracy of power pattern over basis function count for both the PPDM and EFDM. The PPDM accuracy appears to saturate around {\color{black}16 basis functions} and an accuracy of about $-40$~dB, requiring a substantial increase in the number of basis functions for further improvement.
 The EFDM performs about 10~dB better in accuracy for a comparable number of basis functions, getting down to nearly $-50$~dB before the saturation. 

 Both decomposition methods show a significant turning point at which the addition of more basis functions yields little to no improvement in the recovered accuracy. This is expected since SVD produces ordered basis functions with higher order functions contributing less information and eventually reducing to numerical noise. 
 Figure~\ref{fig: first 24 SVD vh} gives a good demonstration of the reducing structure present within the higher order angular basis functions.

\begin{figure}
        \centering
        \includegraphics[width=\linewidth]{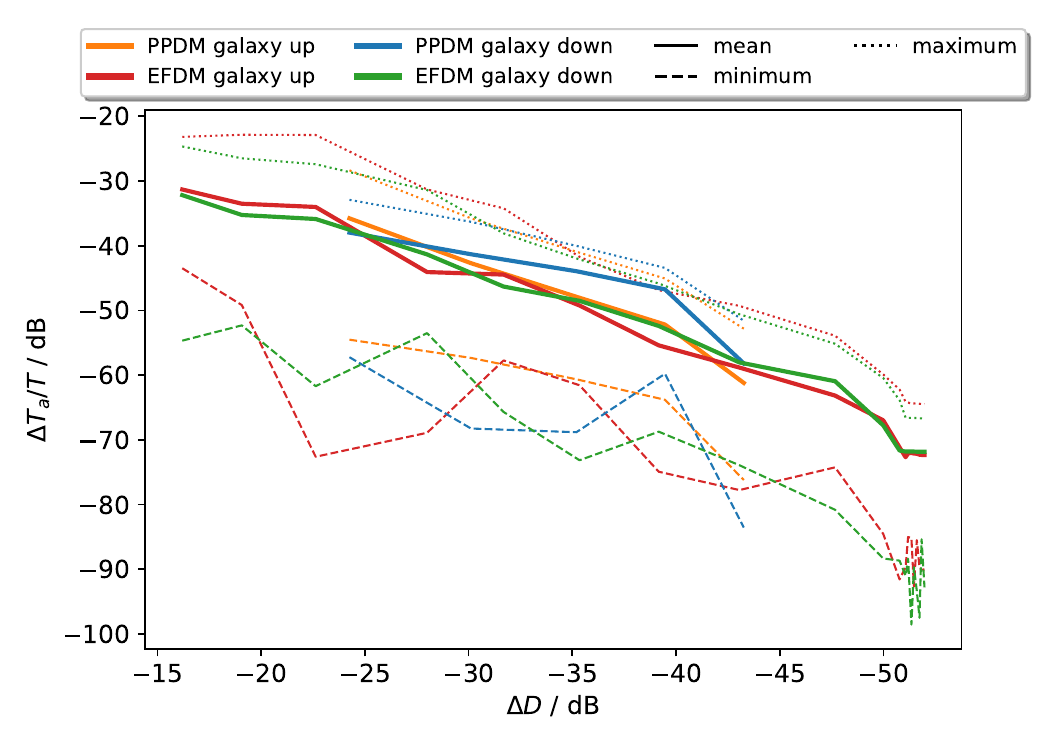}
        \caption{$\Delta D$ \eqref{eqn: delta D} vs $ \Delta T_\text{a}/T$ \eqref{eqn: DT} evolution for galaxy up and galaxy down antenna temperatures for both the PPDM and EFDM pattern reconstructions. The error in estimate of the antenna temperature is lower in both cases for the galaxy being down, and all curves show an increase in antenna temperature accuracy as the beam difference is reduced (until a point of numerical error at better than $-50$~dB). {\color{black}The minimum, maximum mean values of $\Delta T_a / T$ over frequency for each $\Delta D$ evaluation are shown.}}
        \label{fig: D as AT}
    \end{figure}

For the application to a global 21~cm experiment it is important to consider the effect of uncertainty in a power pattern on the expected observed antenna temperature. Figure~\ref{fig: D as AT} shows $\langle \Delta T_\text{a}/T \rangle_\nu$ for both the PPDM and EFDM reconstructions in addition to their range over frequency for varying values of $\Delta D$. These temperatures are calculated as single time snapshots with the galaxy above the horizon (galaxy up) and the galaxy obscured below the horizon (galaxy down). As might be expected, the uncertainty in $T_\text{a}$ reduces as the difference between power  patterns is reduced. 
Note also, at the high accuracy power pattern reconstructions in Figure~\ref{fig: D as AT}, the effect of numerical precision and map resolution on the calculation of $T_\text{a}$ is seen once an accuracy of $-50$~dB is reached in $\Delta D$, and $\Delta T_\text{a}/T$ falls below $-60$~dB. Here, the interpolation between degree gridded $D(\nu,\Omega)$ and \textit{HEALpix} map plays a significant role in the difference between calculations, providing an upper limit on the usefulness of increased accuracy in $\Delta D$ with this computational setup.

\begin{figure}
    \centering
    \includegraphics[width = \linewidth,trim={0 1cm 0 3cm},clip]{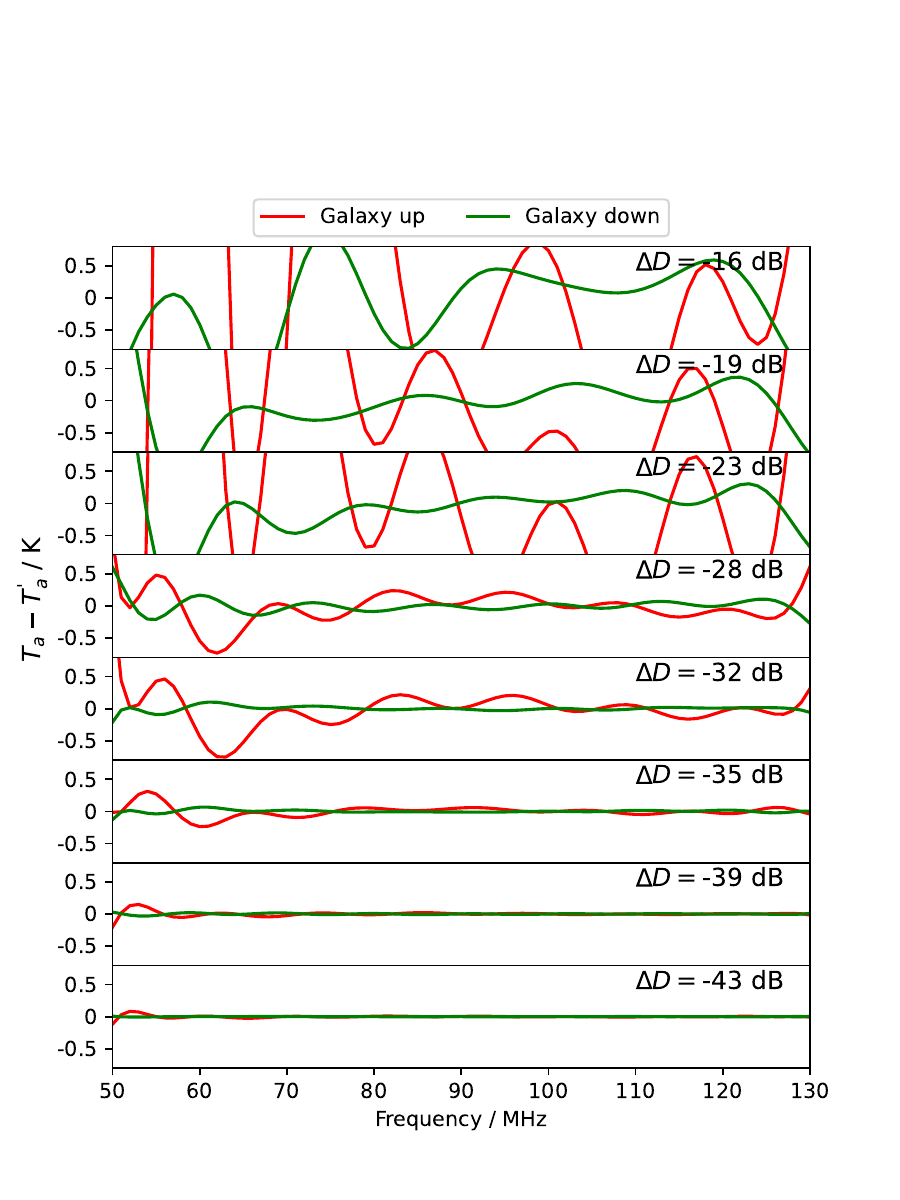}
    \caption{The residual differences in antenna temperature for a selection of accuracies using the EFDM. The residual shows oscillatory structure, which would be able to disrupt a Global 21~cm signal measurement, when at a similar magnitude to the signal. Once $\Delta D < \text{-35 dB}$ the residual seen is below the 200~mK expected of the global 21cm signal.}
    \label{fig: Antenna_temperature_Differences}
\end{figure}

The residual $T_\text{a}$ features generated between different power patterns generally have oscillatory nature which is disruptive to a global 21~cm experiment due to the ability of these troughs to resemble a global 21~cm signal. Examples of the $T_\text{a}$ residuals for a range of different EFDM variations with a $\Delta D$ varying between -16 dB and $-43$ dB are shown in Figure~\ref{fig: Antenna_temperature_Differences}. Lower agreement patterns result in residual structure significantly above 500~mK, while an agreement at a level of roughly $\Delta D = -35\text{ dB}$ is required to reduce the residual structure below 1~K peak-to-peak. Due to the power law structure of the sky temperature larger residual structure is seen at low frequencies, with a first peak still at a magnitude around 100~mK with $\Delta D = \text{-43 dB}$ for the galaxy up antenna temperature calculation.  These residual structures are likely to disrupt a global 21~cm signal experiment when at a level above 200~mK, as is seen for values of dD worse then $-35$~dB.

Table~\ref{tab: abs Ta difference} shows $\Delta D$ and $\langle \Delta T_\text{a}/T\rangle_\nu$ for the SVD decomposition in addition to the maximum absolute uncertainty in antenna temperature for both galaxy up and galaxy down observations. At $\Delta D = -35\text{ dB}$ we see that the maximum absolute difference in $T_\text{a}$ is around the expected level of the global 21~cm signal, below 500~mK, for the galaxy down case. For a $\Delta D = -43\text{ dB}$ the difference in temperatures is below 50~mK, for both galaxy positions, well below the expected level of the global 21~cm signal. As a preliminary calculation this suggests that a knowledge of the antenna beam to a level greater than $\Delta D  = -35$~dB is required for a successful detection of the global 21~cm signal. 
 
\begin{table}
	\centering
	\caption{Average antenna temperature difference, $\langle\Delta T_\text{a}/T\rangle_\nu$ \eqref{eqn: DT}, and absolute antenna temperature difference, $T_\text{a}$ for different levels of $\Delta D$. Increasing accuracy in the power pattern corresponding to decreasing uncertainty in the antenna temperature.}
	\label{tab: abs Ta difference}
	\begin{tabular}{lcccr} 
		\hline
         {} & \multicolumn{2}{c}{Galaxy up} & \multicolumn{2}{c}{Galaxy down}\\
		  $\Delta D$ & $\langle\Delta T_\text{a}/T\rangle_\nu$ & $\text{max} |T_\text{a}-T'_\text{a}|$ & $\langle\Delta T_\text{a}/T\rangle_\nu$ & $\text{max} |T_\text{a}-T'_\text{a}|$ \\
		\hline
            \multicolumn{5}{c}{PPDM} \\
            \hline
		   -24 dB & -35.8 dB & 13300 mK & -38.0 dB & 1470 mK\\
          -30 dB & -42.7 dB & 2190 mK  & -41.3 dB & 539 mK \\
          -35 dB & -47.9 dB & 581 mK   & -43.9 dB & 306 mK \\
          -39 dB & -52.2 dB & 296 mK   & -46.7 dB & 119 mK \\
          -43 dB & -61.2 dB & 47 mK    & -58.2 dB & 21 mK  \\
		\hline
        \multicolumn{5}{c}{EFDM} \\
        \hline
        -28 dB & -44.1 dB & 1025 mK & -41.3 dB & 603 mK\\
        -35 dB & -49.2 dB & 312 mK  & -48.3 dB & 130 mK\\
        -39 dB & -55.4 dB & 205 mK  & -52.4 dB & 32 mK \\
        -43 dB & -58.8 dB & 117 mK  & -58.0 dB & 8 mK  \\
        -50 dB & -67.0 dB & 9 mK    & -67.8 dB & 1 mK  \\
        \hline
	\end{tabular}
\end{table}

\section{Quantifying acceptable power pattern uncertainty within a global 21~cm experiment} \label{sec: Errors}

To examine the impact of varying levels of $\Delta D$ on the ability to recover the 21~cm signal when using the REACH pipeline, a number of the constructed power patterns from the previous Section~\ref{sec: SVD} was used to attempt to make a detection of a set of representative 21~cm signals. These reconstructed patterns are selected to have a specified $\Delta D$ error level, and both PPDM and EFDM models are used to provide some diversity in the test patterns.
Gaussian signals are used to replicate the shape and depth of a possible global 21~cm signal trough. The Gaussian signal can be described by 
\begin{equation}
    \label{eqn: 21cm gaussian}
    T_{\text{G21}} = -A_{21}e^\frac{-(\nu-\nu_{21})^2}{2\sigma_{21}^2},
\end{equation}
 where $\nu_{21}$ is the centre frequency, $A_{21}$ is the amplitude and $\sigma_{21}$ is the standard deviation of the signal. This form requires fewer parameters to fit than a fully formed signal, allowing for a more efficient evaluation while giving a good representation of performance with a full signal. For the purposes of this work, five different Gaussian global 21~cm like signals are evaluated (described in Table~\ref{tab: pipeline signals}) using 35 sky regions and 48 time bins over a 240~minute observation from LST = 4.64 hour angle. The signals are chosen such that the centre frequency $\nu_{21}$ spans the full bandwidth of the REACH dipole, and the amplitude to be representative of the majority of global 21~cm signal models. The resulting signal fits of the original power pattern is shown in Figure~\ref{fig: exact beam refit plots}.
 
 Only a select number of signals are simulated due to the high computational time to fit each signal with the required set of test power patterns. Running on the Cambridge University CSD3 icelake system it takes $4~-~14$~hours of real time per signal evaluation for each of the tested power patterns. Signals with higher centre frequencies are faster to evaluate due to the greater difference between the magnitude of the inserted signal and the magnitude of the background temperature. 

\begin{table}
	\centering
	\caption{The Gaussian signals used within the REACH mock detection pipeline for evaluation of impact of $\Delta D$ on detection ability. Using centre frequency $\nu_{21}$, amplitude $A_{21}$ and standard deviation $\sigma_{21}$.}
	\label{tab: pipeline signals}
	\begin{tabular}{lccr} 
		\hline
		  Signal & $\nu_{21}$ / MHz & $A_{21}$ / mK & $\sigma_{21}$\\
		\hline
		  No signal & -  &  0  & - \\
        60~MHz & 60 & 150 & 15 \\
        80~MHz & 80 & 150 & 15 \\
        100~MHz & 100 & 150 & 15 \\
        120~MHz & 120 & 150 & 15 \\
		\hline
	\end{tabular}
\end{table}

We generate a base observation data set by observing the sky with $D(\nu,\Omega)$. The pipeline fit is then run using $\Tilde{D}(\nu,\Omega)$ as the assumed antenna beam, fitting either for a Gaussian signal and foregrounds or for just the foregrounds. Each fit results in the Bayesian evidence, $\log \mathcal{Z}$, or confidence of fit, and preferred parameter values for the signal. Using these values a confidence of detection, $\Delta \log \mathcal{Z}$ \eqref{eqn: DLZ}, and accuracy of detection, RMSE \eqref{eqn: RMSE}, are found and for each refitting.
    \begin{figure}
        \centering
        \includegraphics[width=\linewidth]{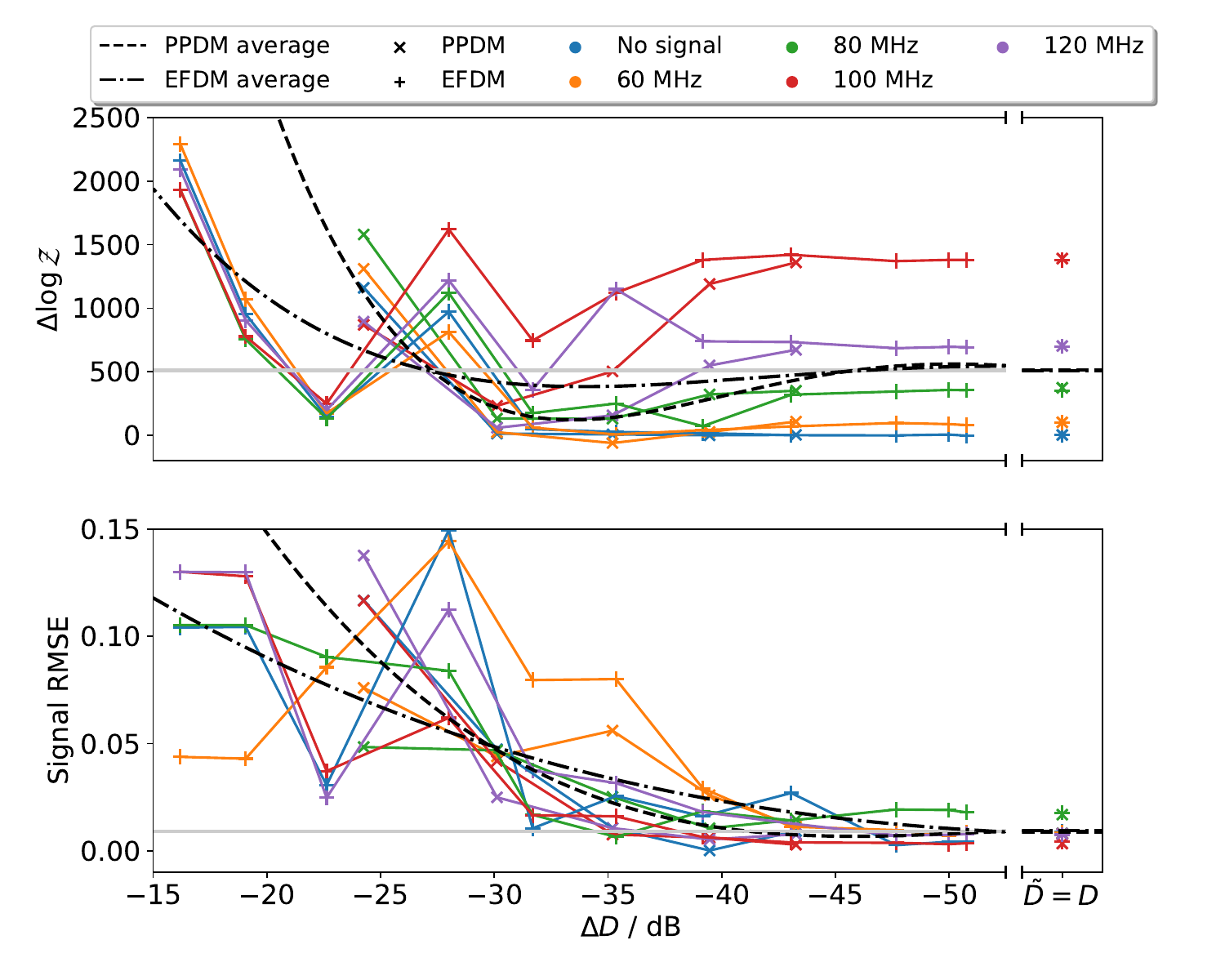}
        \caption{The $\Delta \log \mathcal{Z}$ \eqref{eqn: DLZ} and RMSE \eqref{eqn: RMSE} for REACH pipeline runs for a variety of different $\Delta D$ values. Showing confident but incorrect solutions at high uncertainty in power pattern and increased accuracy and consistent confidence in fit for lower uncertainty in power pattern accuracy.}
        \label{fig: combined dlz vs rms}
    \end{figure}
    \begin{figure}
        \centering
        \includegraphics[width = \linewidth]{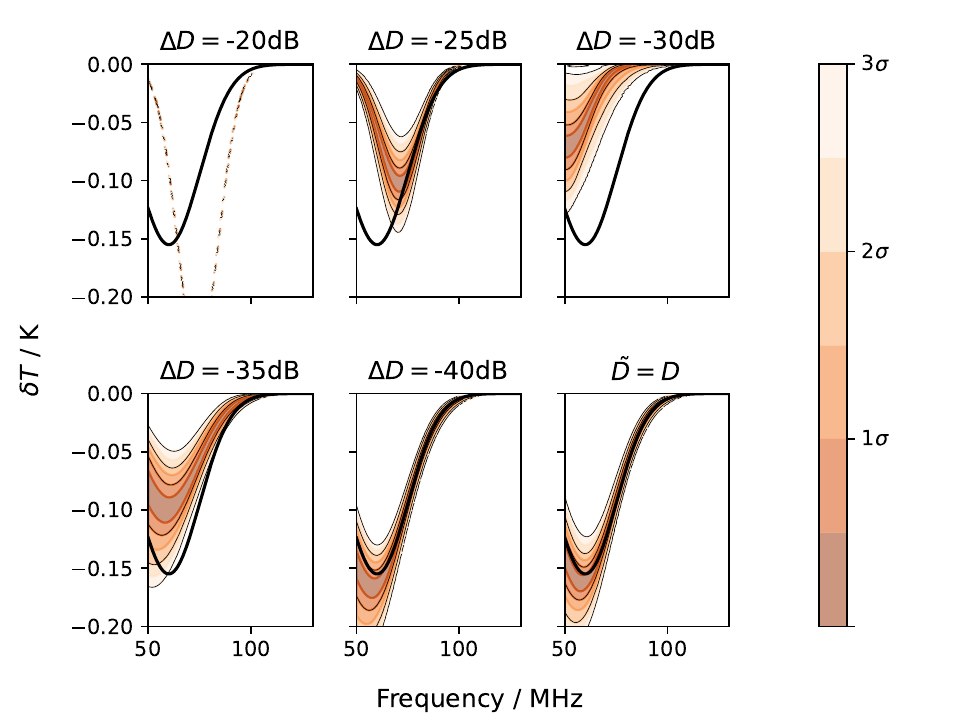}
        \caption{{\color{black}The reconstructed fits by the REACH pipeline with a selection of uncertainties in the power pattern used for data generation and refitting. The injected Gaussian signal at 60~MHz is shown in black. In agreement with figure \ref{fig: combined dlz vs rms} significant biases are shown at low power pattern agreement.}}
        \label{fig: 60_subplots}
    \end{figure}
    The $\Delta \log \mathcal{Z}$ and RMSE differences for the selection of $\Delta D$ value beams are shown in Figure~\ref{fig: combined dlz vs rms}, with each tested against the five signals in Table~\ref{tab: pipeline signals}. At low accuracy, around $\Delta D > -25$~dB, it is seen that the both the RMSE and $\Delta \log \mathcal{Z}$ for the given signals are high. This is likely due to the large residual difference between the two power  patterns acting to generate a significant systematic uncertainty within the data, which is then fit as a signal. As $D(\nu,\Omega)$ and $\Tilde{D}(\nu,\Omega)$ become more similar, it is seen that the RMSE generally decreases to eventually arrive around the value found when $\Tilde{D}(\nu,\Omega) = D(\nu,\Omega)$. This is expected as the reducing difference in antenna patterns will result in smaller residuals generated in the antenna temperature which are less likely to obscure the possible detection of a global 21~cm signal. 
    
    In contrast, $\Delta \log \mathcal{Z}$ first reduces significantly until flattening to approximately 0 between $\Delta D =-30\text{ dB}$ and $\Delta D = -35\text{ dB}$. At this point the differences in the antenna beam makes it difficult to distinguish between the absence and presence of a signal. This trend towards a minimum in the $\Delta \log \mathcal{Z}$ between $-30$~dB and $-35$~dB is seen for all the centre frequencies of the signals. As seen in Table~\ref{tab: abs Ta difference}, this level of difference between antenna beam patterns is where antenna temperature uncertainty begins to approach the magnitude of the global 21~cm signal, at around 300~mK to 500~mK. For $\Delta D < -35\text{ dB}$ we again see an increase in confidence of fit as the systematic residuals generated from the antenna pattern error are smaller than the magnitude of the inserted Gaussian signal. 
    
    Here we see that $\Delta \log \mathcal{Z}$ is highest for signals at higher $\nu_{21}$, corresponding to a lower foreground temperature, as well as a signal entirely contained within the available bandwidth of the observation between 50~MHz and 130~MHz. Therefore, the best and most confident reconstructions are for a signal at 100~MHz, while the signal at 60~MHz remains at lower confidence even when $\Tilde{D}(\nu,\Omega) = D(\nu,\Omega)$. {\color{black}This observation agrees with the hypothesis that a lower frequency signal will present a greater challenge to detection due to the power law nature of the synchrotron foreground.}

    Finally, the $\Delta \log \mathcal{Z}$ seen for the null signal does not rise again for values lower than $\Delta D = -40\text{ dB}$. This is expected given a correct detection in this case is with no signal, so one would expect to have the same evidence when refitting for a signal and with only foregrounds. At this level of $\Delta D$ the residual antenna temperature features introduced into the data are lower than those which might be detected for a global 21~cm signal. This threshold is important for identification, as this is the level of accuracy at which an antenna beam pattern is required to be known to prevent a false detection in the case of no signal being present within the data.

    {\color{black}An example evolution in the reconstructed signals against differing $\Delta D$ is shown in Figure~\ref{fig: 60_subplots}. Here it is seen that for poor agreement between power patterns significant bias is seen within the reconstructed signals, while at $\Delta D = -40\text{ dB}$ a result comparable to using the same pattern for data generation and analysis is achieved.}

    \begin{figure}
        \centering
        \includegraphics[width = \linewidth]{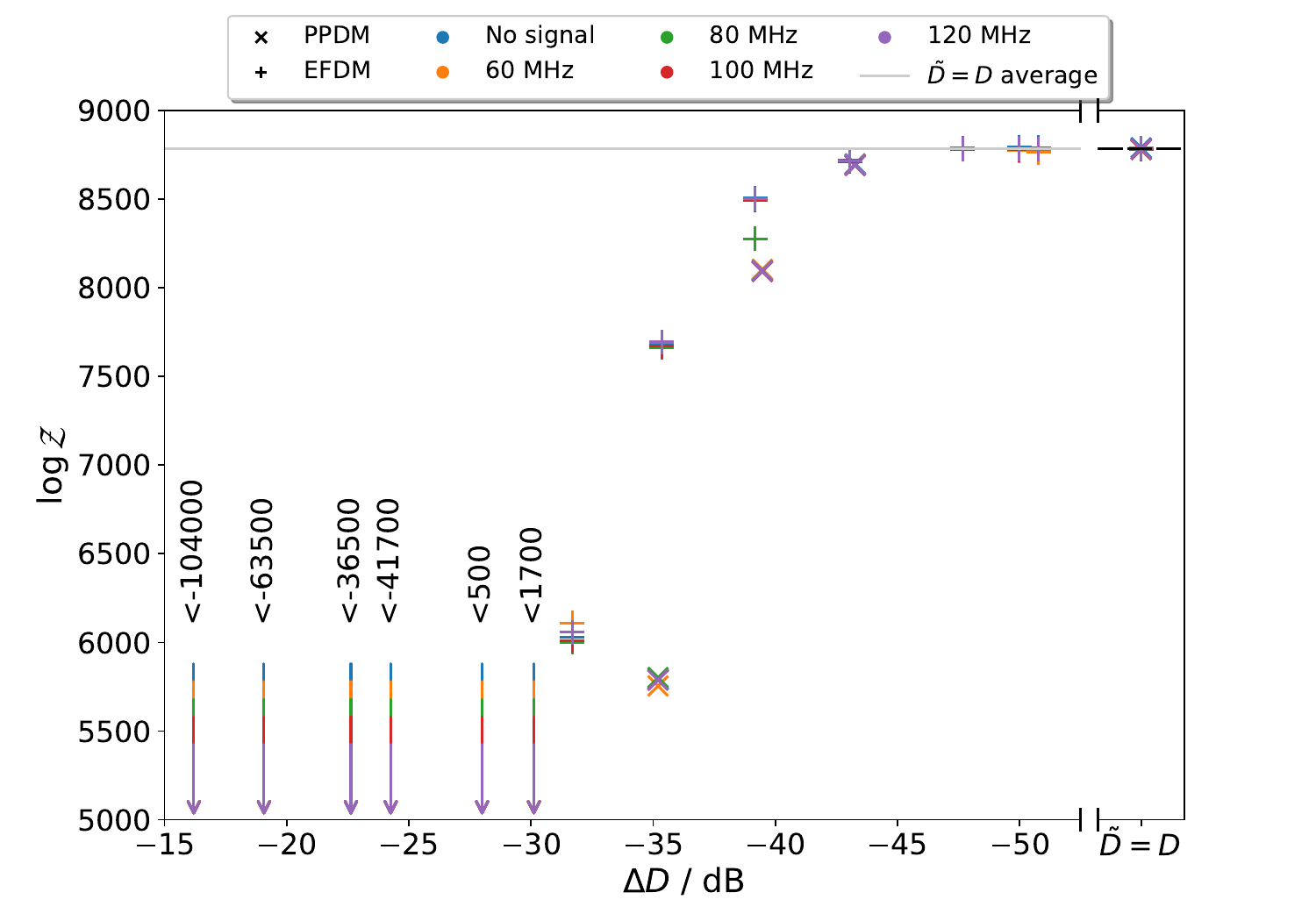}
        \caption{The $\log \mathcal{Z}$ associated with the fits shown in Figure~\ref{fig: combined dlz vs rms}. The evolution towards higher $\log \mathcal{Z}$ is expected for this type of Bayesian analysis, and reinforces that more accurate power pattern models produce higher confidence in results.}
        \label{fig: LogZ}
    \end{figure}
    For a confident detection to be possible within the current REACH pipeline framework, Figure~\ref{fig: combined dlz vs rms}, suggests that a difference between the observing beam, $D(\nu,\Omega)$, and that used for the data analysis, $\Tilde{D}(\nu,\Omega)$, of better than $-30$~dB accuracy is acceptable to avoid the detection of an incorrect signal. At around $-35$~dB accuracy it is possible that a detection can be made, at accuracy better than $-40$~dB the evidence and RMSE are at approximately the same level as if the same beam pattern is used for data generation and observation. 

\section{Power pattern difference due to antenna geometry} \label{sec: Power pattern difference}

\begin{figure}
    \centering
    \includegraphics[width = 0.8\linewidth]{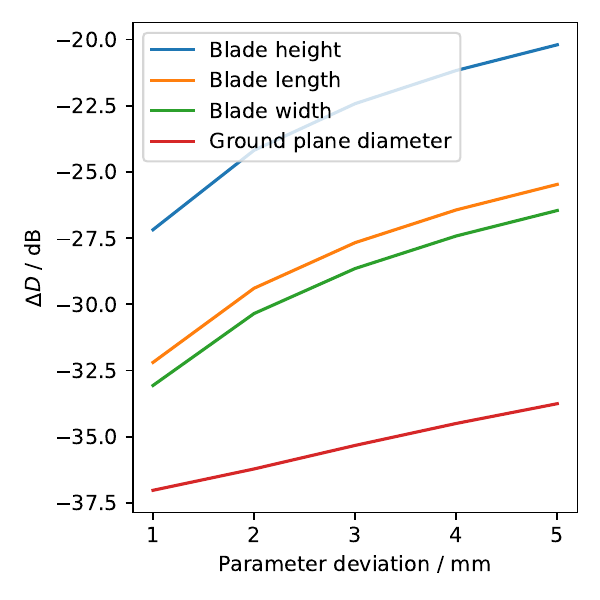}
    \caption{\color{black}The $\Delta D$ between the modified REACH dipole with height 1000~mm compared to antennas of varying height, blade dimension and ground plane size. A match better than -20 dB requires a difference in height less than 5~mm from 1000~mm. The tolerance in blade length is larger, with $\Delta D$ of better than -30 dB being achieved for variations under 3 mm. $\Delta D$ remains below -35~dB for variations in the ground pane diameter up to the 5~mm shown here.}
    \label{fig: height vs dD}
\end{figure}

{\color{black}In Section \ref{sec: Uncertianty}, we listed a series of possible causes of uncertainty within an antenna power pattern and in Sections \ref{sec: directivity accuracy} and \ref{sec: Errors} we have identified a limit of required accuracy for knowledge of the antenna beam pattern. To present this change in the context of the construction of an antenna, here we include a brief exploration of the power pattern uncertainty generated due to small changes in the physical dimension of the antenna.

A selection of simulations were carried out varying the height or blade length of the modified REACH dipole, with $\Delta D$ then calculated between each pattern relative to the base antenna power pattern. These results are shown in Figure \ref{fig: height vs dD}. There is seen a strong relationship between antenna height and $\Delta D$, with a difference of 5~mm from the base height of 1000~mm resulting in a $\Delta D = -20\text{ dB}$, to achieve a match better than $-25$~dB a height difference of less than 1~mm is required. The tolerance in the x-axis direction blade length is shown to be less sensitive to variation, with $\Delta D$ better than -30~dB being achieved with variations less than 3~mm. These parameters are expected to be two of the most sensitive, and they highlight the importance of a physically accurate model for the antenna to be used.}

\color{black}
 
\section{Conclusions and possible compensation} \label{sec: Compensation}
    {\color{black}Through this paper we have considered the possible effect of an inaccurately known antenna power pattern upon the ability of a REACH-like global 21~cm experiment to perform an accurate detection.

    Due to the inability to completely measure and simulate the full environment of the antenna perfectly, it will be unavoidable that uncertainty is present within an antenna beam used within data analysis; for REACH or any similar global 21~cm experiment. This uncertainty can then be expected to have an impact on the possible level of confidence for a possible detection. It is shown that variations in the antenna height of order 1~mm are able to generate sufficient changes to the power pattern to disrupt a detection of the global 21~cm signal.

    For the work presented in this paper, a simplified version of the REACH hexagonal blade dipole antenna was used as an example reference antenna power pattern. A selection of perturbed power patterns was constructed by performing singular value decomposition on the electric field and power pattern of the reference power pattern. By employing two different sets of basis functions, some diversity was given to the set of basis functions and in turn to the perturbed beams. It was shown that the difference between the perturbed and reference power patterns requires $\Delta D < -35$~dB for the resulting antenna temperature difference to be of the same order as the magnitude of the larger plausible models for the global 21~cm signal, around 300~mK.

To further assess the impact of power pattern uncertainties upon a real experiment, we employ the REACH mock data analysis pipeline to attempt the detection of a selection of Gaussian models of the global 21~cm signals. The Bayesian evidence given by the pipeline is used to determine the confidence in signal detection, by comparing the evidence for fits with and without a signal present. This is also compared to the RMSE of the injected and reconstructed signal. We find that for a power pattern difference worse than $-35$~dB the experiment would be susceptible to making false signal detection. At $\Delta D = -35$~dB the confidence level of the detection lessens, but the signal is more accurately recovered. With $\Delta D = -40$~dB we find that confident and accurate detections are consistently performed. This suggests, without additional compensation methods in place, that knowledge of the power pattern used within the REACH pipeline should be at least at this level compared to the observing power pattern.

    Due to the likely presence of errors at this level within a global 21~cm experiment, the use of some form of compensation and precaution is required to aid the confidence in a possible detection. We shall now discuss a selection of these.

    Firstly, ensuring that the computational simulation of the antenna beam is as accurate to reality as possible. Although this is highly time and computationally intensive, one should incorporate as much physical detail as is known into the simulation model. This includes measurements of small deviations in the ground plane and other structures nearby to the antenna, an accurate soil and ground topography model, antenna feed details, etc. Similarly, using and comparing a selection of CEM software and methods will also allow identification of possible discrepancies.

    To accompany the use of different software and CEM methods, an approach could be taken where several likely candidate patterns are used within the detection pipeline framework. Figure~\ref{fig: LogZ} demonstrates that the calculated $\log \mathcal{Z}$ for the fits of different beam uncertainties increases as the accuracy of the fit increases. So, if a selection of likely candidate power  patterns are identified these could all then be used independently and the eventual highest $\log \mathcal{Z}$ taken, which will show preference towards the most accurate candidate beam.

    A more advanced method of compensation is to incorporate variability of the power pattern within the data analysis process itself. Although it is possible to reconstruct any power pattern using analytical mathematical functions, such as spherical harmonics, these will often require many thousands of coefficients as seen in Section~\ref{sec: sphericals recon} to achieve the required accuracy. The huge amount of additional parameters this causes makes a direct incorporation into the REACH pipeline computationally untenable. Therefore, more informed basis functions are required to be used, such as those based upon an SVD discussed in Section~\ref{sec: Directivity reconstruction}. The use of the SVD-based numerical basis functions of the pattern allows for a realistic variation in the power pattern to be achieved while dramatically reducing the number of fitting parameters, requiring tens of distinct basis patterns rather than many hundreds. Incorporation of variation of these basis functions (extracted from CEM models where the geometry of the antenna is also varied along with frequency) into the REACH pipeline would allow for the dynamic fitting of a power pattern at the same time as the remaining sky and signal parameters, thereby reducing the eventual uncertainty between observing and analysis power patterns. Work is ongoing to incorporate a beam fitting method within the REACH pipeline, which while promising, requires a further reduction in the number of fit parameters to be efficient.


\section*{Acknowledgements}
JC is supported by the Science and Technology Facilities Council (STFC) through grant number ST/X00239X/1. CP and DdV are supported in part by the South African Radio Astronomy Observatory, which is a facility of the National Research Foundation, an agency of the Department of Science and Technology (Grant Number: 75322). EdLA is supported by the STFC ERF grant ST/V004425/1. Some of the results in this paper have been derived using the Healpy and HEALPix packages. The authors would like to thank Dominic Anstey for his work developing the REACH Bayesian pipeline and advice.


\section*{Data Availability}
The data that supported the findings of this article will be shared on reasonable request to the corresponding author.
 



\bibliographystyle{mnras}
\bibliography{main} 




\appendix




\bsp	
\label{lastpage}
\end{document}